# Investigating students' strengths and difficulties in Quantum Computing.

Tunde Kushimo
*Department of Physics and Astronomy, Southern Methodist University, Dallas, TX*

Beth Thacker
*Department of Physics and Astronomy, Texas Tech University, MS 41051, Lubbock, TX*

**Abstract:** Quantum Computing is an exciting field that draws from information theory, computer science, mathematics, and quantum physics to process information in fundamentally new ways. There is an ongoing race to develop practical/reliable quantum computers and increase the quantum workforce. This needs to be accompanied by the development of quantum computing programs, courses and curricula coupled with the development and adoption of evidence-based materials and pedagogies to support the education of the next generation of quantum information scientists and workforce. At our university, we introduced an introductory course in Quantum Computing to undergraduate students and conducted a case study to investigate the strengths and difficulties of these students in quantum computing after taking the introductory course. Our goal is to contribute to the improvement of quantum computing education while understanding the topics that the students find easy to comprehend and topics that are difficult to comprehend. We conducted a series of interviews to identify these strengths and difficulties. This investigation is essential to help train the next generation of quantum scientists and researchers because it gives information and direction to quantum computing education, and it also helps us to develop research-based teaching and learning materials. We report on the results of these interviews and our initial work on the development of evidence-based materials for teaching an introductory course in Quantum Computing.

# I. INTRODUCTION

Quantum Information Science and Quantum Computing are exciting fields that draw from information theory, computer science, mathematics, and quantum physics to process information in fundamentally new ways. Quantum mechanics being a more general model of physics than classical mechanics gives rise to a more general model of computing. Quantum mechanics describes certain puzzling behaviors of matter and light, which are now being used in Quantum Computing. Some major properties of quantum systems that distinguish them from classical systems are their probabilistic nature, superposition, entanglement, interference, and measurement. These collective properties of quantum systems can be harnessed to acquire, encode, store, transfer, and manipulate information and perform computation. It is believed that quantum computing will redefine computing and information science, proffering solutions to problems that seem classically impossible to solve while also opening opportunity for more advanced computational works and studies. Thus, there is a race to develop practical quantum computers. To achieve this, more people need to be trained and educated in the fields of quantum computing and quantum information science.

A tremendous number of resources, both governmental and commercial, are being invested in quantum computing technologies and research to realize practical quantum computers. In December 2018, the President of the United States signed into law the National Quantum Initiative Act (NQI), creating a multi-agency program, including the National Institute of Standards and Technology (NIST), the National Science Foundation (NSF), and the Department of Energy (DOE) [1]. This initiative promotes research in QC and QIS, in the training and *education* of students to enter the quantum workforce, and the retraining of the existing workforce. This includes forming several Quantum Information and Science (QIS) Centers to support research *and education* in Quantum Computing (QC) and QIS. Again, QIS continues to gain more attention locally and globally as the President of the United States signed an executive order to advance quantum technologies in 2022 [2]. This is not specific to the United States, but a global issue as many other governments are investing into QC and QIS related research [3-5].

QC research and education is gaining lots of attention from governments, academics, industries, researchers, engineers, and investors. The number of introductory courses in quantum computing at the undergraduate and graduate levels is increasing across the country; it has even extended to high schools. In the four years that the NQI Act has been signed into law, Quantum Computing Report [6] recorded that there are approximately seventy Universities with Quantum Computing research groups offering some form of Ph.D. degrees in this field. In addition, many other universities offer elective courses, certificate courses, and master's degrees in Quantum Computing. There are also many online courses on Quantum Computing organized by universities and industry experts. Several educational courses, and resources on Quantum Computing have been developed to meet the increasing demand for a growing quantum workforce. Some freely available resources include the IBM Qiskit textbook [7], Quirk [8], D-Wave leap community [9], Azure Quantum [10], Qubit by Qubit [11], and many more. Apart from the Ph.D. and master's degree courses in Quantum Computing that have defined program plans, the majority of the elective courses, certificate courses, and online courses are introductory courses often with a requirement of a basic knowledge of linear algebra. Even though these programs and courses are doing their best to educate the present and future quantum workforce, we hypothesize that course design would benefit from Physics Education and Computer Science Education Research.

An introductory course in Quantum Computing covers the basic knowledge that students will need about the emerging field to understand and pursue further education and careers in various research areas of quantum computing and quantum information science. Generally, this introductory course introduces students to the fundamental topics in Quantum Computing while also training them in the language of the emerging field. We hypothesized that students would gain some strengths and have some difficulties learning about Quantum Computing for the first time. These students' strengths and difficulties are the concern of our investigation.

In this study, our research question is: What are the students' strengths and difficulties after taking an introductory course in quantum computing?

This investigation is essential to help train the next generation of quantum scientists and researchers because it gives information and direction to quantum computing education while also informing the development of evidence-based teaching and learning materials.

## II. INTERVIEW GROUP

The interview group for this qualitative study was students that had taken an introductory course in quantum computing at the undergraduate level. To recruit interviewees, we addressed a group of quantum computing students explaining our research plan and their role in it. Five students that took the Introduction to Quantum Computing course the semester before this research was conducted volunteered [Table 1]. A participation incentive of $15 per interview session was given to the students. All of the students were Physics majors. We did not have Computer Science majors in this class. One of the students we interviewed had taken a Computer Science course on logic tables and gates prior to taking the introductory course on quantum computing while others had no knowledge of it before taking the Introduction to Quantum Computing course. All the undergraduate students had some form of introductory course on quantum mechanics, either before or while taking the introduction to quantum computing course. None of the students had had any prior education in quantum computing before taking the Introduction to Quantum Computing course.

| Students | Education level | Major | Basic knowledge of linear algebra | Previously had a quantum mechanics course | Prior knowledge of classical logic gates and tables |
| --- | --- | --- | --- | --- | --- |
| A | Senior | Physics | Yes | Yes | Yes |
| B | Senior | Physics | Yes | Yes | No |
| C | Senior | Physics | Yes | Yes | No |
| D | Senior | Physics | Yes | Yes | No |
| E | Junior | Physics | Yes | Had it concurrently with the QC course | No |

Table 1. Information about the interviewees

## III. CLASS FORMAT AND LEARNING CONTENT

The undergraduate course was taught to undergraduate students using a student-centered, interactive learning approach in a semi-flipped classroom environment. The students were encouraged to read ahead and come to class prepared to discuss, work problems and answer questions in small groups or sometimes as a class.

The course was developed with the goals of helping students to understand two-level quantum systems and the applicability of the basic postulates of quantum mechanics in quantum computing.

As in most introductory Quantum Computing courses, the following topics were covered:
- Two-level systems
- Superposition and Entanglement
- Measurement and Probability
- Classical and Quantum bits
- Quantum gates
- Quantum circuits
- No-cloning theorem
- Quantum parallelism
- Quantum Teleportation
- Superdense coding
- Quantum Key Distribution
- Quantum Algorithms: Deutsch, Deustch-Josza, Berstein-Vazirani, Simon's, Grover's, Quantum Fourier Transform

We stopped short of teaching Shor's algorithm due to a lack of time at the end of the semester.

We chose to teach the basic quantum concepts from the first two chapters of a quantum text [12] and then used the online Qiskit textbook [7] and IBM quantum platform as the basis for the rest of the course. We supplemented those materials with other available texts and online resources [13-18].

**IV. METHODOLOGY**

In qualitative research methods like case study, semi-structured interviews are appropriate tools used to extract information and establish deeper understanding of topics. This method is a mixture of dialogue and examination, guided by flexible interview questions and probes used to collect data about the research topic [19 - 21]. Data collection for this research was done using the semi-structured interview method. We chose this method because it is one of the most effective ways to understand how our interviewees interpret and answer our questions, their thought patterns, and their difficulties as they answer the questions. We took time to go through every topic in the syllabus and set some questions from each topic before deciding the final questions to include in the interview process. The final set of interview questions was focused mostly on the beginning and middle of the course and did not include questions on the different Quantum Algorithms taught. The interview questions were a predetermined series of conceptual and exam-like questions with structured probes attached to every question [22] [see an example question with probes in Fig. 1]. The structured probes were intentionally designed to query students' strengths and difficulties. Even though the questions were structured, the interview sessions were flexible enough to be discussion-driven, and the interviewees were encouraged to answer the questions using the think-aloud method.

We conducted two one-hour long sets of interviews for each student. The first interview set was over the first third of the syllabus with seven main questions, while the second interview was over the second third of the course (up to Quantum Algorithms), with five main questions. Since our focus was on the students' thinking skills and their ability to apply their learning, we provided them with markers and whiteboards to write on while they thought out loud as they answered the questions. This allowed us to pay close attention to their thought processes for further probing. We also provided the students with prompts during the interviews when we deemed it necessary to drive the needed discussions. We made video recordings of every interview session and took notes while observing our interviewees during the interview sessions [23]. These notes and video clips were revisited for each of the interviewees, and we analyzed them using the thematic analysis method.

We repeatedly went over each data transcript, looking for clues about students' strengths, knowledge gaps, and difficulties as they answered questions and during discussions. For all the students, patterns and clues suggested what they were comfortable with and confident about and what they struggled and had difficulties with. Such clues as students hesitating, taking longer thinking time than expected, answering a question incorrectly or outrightly saying that they could not explain how they arrived at an answer suggested struggles and difficulties.

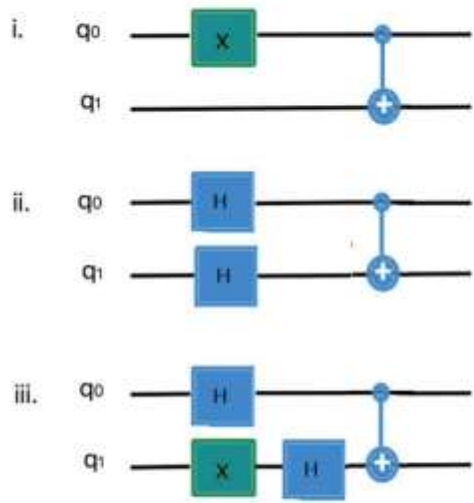

FIG. 1. An Example question with structured probes

In contrast, clues like detailed and step-by-step explanations suggested strengths. We took note of such patterns and drew meaningful results from them.

With the thematic analysis method [24 - 27], we were able to search, identify, analyze, organize, define, and report students' strengths and difficulties in the interview transcripts. We identified and recorded the main patterns and ideas via the following steps of the thematic analysis method:

**i. *Familiarization with the interview transcripts***: We immersed ourselves with the data to get familiar with the details of the transcripts. We did this by watching and listening to the video recordings repeatedly, looking for patterns and ideas that suggested strengths and difficulties.

**ii. *Searching and generating patterns***: From repeatedly listening to the interview transcripts, we analyzed and reflected on the data. We moved from unstructured data to the development of themes, ideas, and patterns, focusing on important sections of the transcripts/data. We began sorting and collating all the potentially relevant themes that were related to the study. We categorized specific solutions and statements from the students into patterns that represent our study.

***iii Reviewing patterns***: At this stage, we revisited all the individual observed themes and patterns if they were coherent. We checked to confirm that the patterns recorded as strengths and difficulties were common to at least 60% (N=3, three-fifths of the students) of the interviewees' population.

**iv. *Labelling patterns***: We labelled and categorized the patterns. Some of the patterns collapsed into each other, so they were recorded under the same categories. This stage of the analysis process was repeated as we kept modifying and refining the categorization.

It is necessary to mention that there were strengths and difficulties that were recorded among one to two students (below 60% of the interviewee's population). Such strengths and difficulties were not reported in our final step of the thematic analysis explained above. We recorded and reported strengths and difficulties common to 60% (N = 3) of our interviewee's population.

## V. RESULTS

We identified students' strengths and difficulties based on their answers to the interview questions and how they explained their approaches to the questions using the thematic analysis method. The following points discuss the results of students' strengths and difficulties in learning in an introductory course in quantum computing from the analysis of the recorded data. We summarize the results in Table 2.

| Strengths | Difficulties |
|---|---|
| Measurement | Dirac operations |
| Matrix operations | Operations of the rotation operators, and Toffoli gate |
| Bloch sphere | Phase kickback |
| Operations of the Pauli X and Z, Hadamard, and CNOT gates | Notations across various texts and online resources. |
| Interpreting quantum circuit diagrams | |

Table 2. Summary of results

### A. Strengths

The following are the reported strengths common to 60% and above of our interviewed students.

- *Measurement:* All the interviewed students confidently explained that the results from measuring qubits are always in the form of zeros and ones with some probability values; it cannot be in a superposition state (Fig. 2). Measurement is an important concept to understand in both quantum mechanics and QC. Quantum states may be in a superposition or in an entangled state, but the act of measurement collapses the state into a basis state.

**Interview excerpt:**
**Situation:** What can you say about superposition states and measurement?
**Response:** *The thing about superposition is that when we measure, we get one state, having a one or zero. We can calculate the probability of it.*

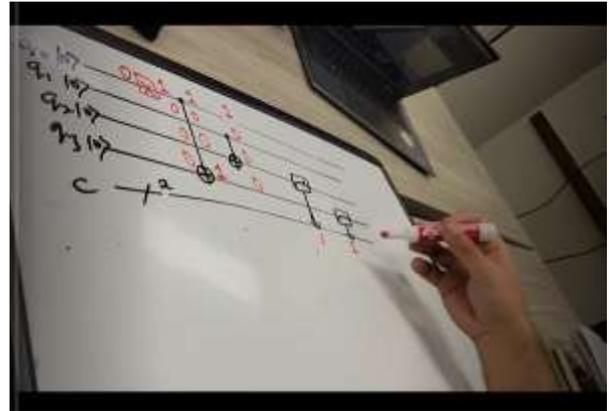

FIG. 2. An example of a student explaining how measurement collapses quantum states

- *Matrix operations:* For almost all the interview questions that involved doing some calculations, many of the students approached it using matrix methods. This was a pattern we observed during analysis. The students were very confident and comfortable using matrix operations for almost all the questions (Fig. 3 shows a student using matrix method in a calculation). They got most of the solutions right, except for questions where applying matrix methods seemed complicated and unnecessarily long. Linear algebra (of which matrix operation is a major part) is a mathematical tool used in QC. Quantum gates and their operations on qubits can be represented and computed in matrix form, understanding how that works is a good step to understanding, writing, and interpreting quantum algorithms.

**Interview excerpt:**
**Situation:** How were you able to know how the gates affect the qubit states?
**Response:** *Whenever I do these calculations, I always work in matrix*
**Another response:** *I like to think about these things in terms of a matrix*

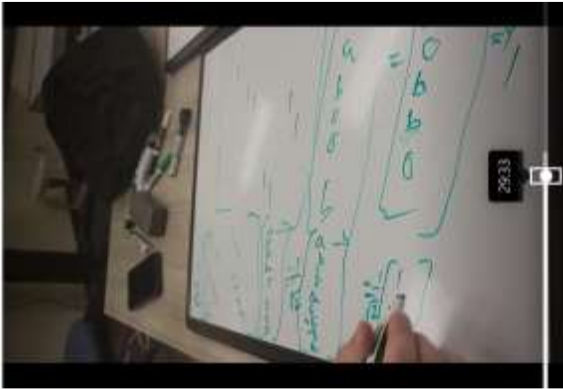
FIG. 3. Screenshot of a student using matrix method in a calculation

- **Bloch sphere:** All the interviewees understood how qubit states are represented and manipulated on the Bloch sphere. They drew, labeled, and explained the z, y, and x-basis states, and several superposition states. (Fig. 4). Bloch sphere is a good way to visualize single qubits (various basis and superposition states) and how their states are manipulated by the actions of single-qubit gates. Understanding this helps the students' conceptual understanding of other important concepts like global phase and relative phase.

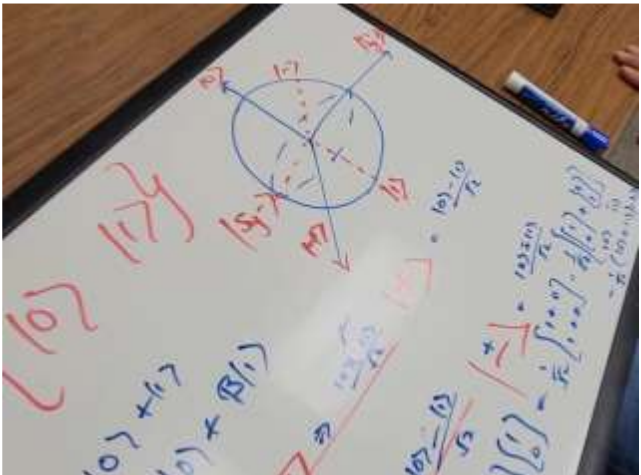
FIG. 4. Screenshot of a Bloch sphere question solved by a student

- **Pauli X and Z, Hadamard, and Controlled-X gates:** According to our analysis, our interviewees were very confident about how to represent and operate these important QC gates. They were able to show and explain the operations of these gates in quantum circuits diagrams, via matrix operations, and on the Bloch sphere. One of the students mentioned during the interview that they gained this strength easily because every example, simulation, and question they worked on had these major gates. (Fig. 5). Such strength as this is very important for both theoretical and research work in QC because these are essential gates in the discussion of universal quantum gates.

**Interview excerpt:**
**Situation:** Can you talk me through the quantum circuit and what each gate does (see circuit figure below)?
**Response:** *When you apply X-gate on zero state, it flips it to one*
**Another response**: *When this CNOT acts upon the state because the control qubit is given in state one, the target qubit will then be a value of one from zero.*

**Situation: How would you put a state in a superposition state?**
**Response:** *This is the Z-basis, but you can change it to the X or Y basis and the easiest way is the Hadamard gate. If you apply Hadamard gate to like zero, it would change to one over square-root two zero state plus one over square-root two one state. So, it is now a combination of both zero and one which also can be written as the plus state in the X-basis.*

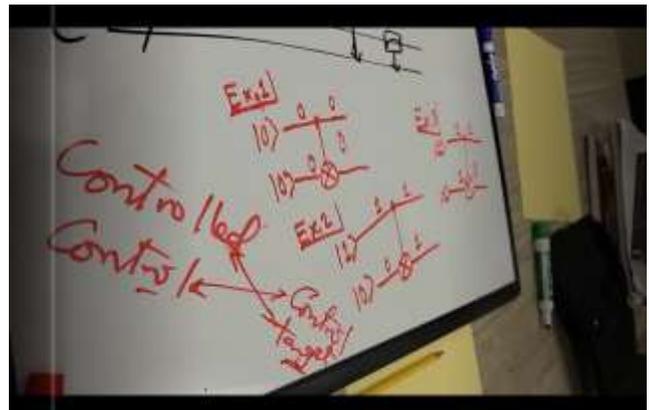
FIG. 5. Screenshot of a student's work showing how the CNOT gate manipulates qubits

- **Interpreting quantum circuit diagrams:** This was one of the major strengths recorded from our analysis. The majority of the students were able to perform step-by-step explanations of circuit diagrams, describing the qubit states before and after the application of gates. They interpreted successfully both single qubit circuits and multiple qubits circuits. (see Fig. 6)

**Interview excerpt:**
**Situation:** How were you able to work through the quantum circuit so easily?
**Response:** *I believe when it comes to quantum circuits, it will be very helpful to keep track of the states before and after each operation, showing how the gates manipulate the qubits per time.*

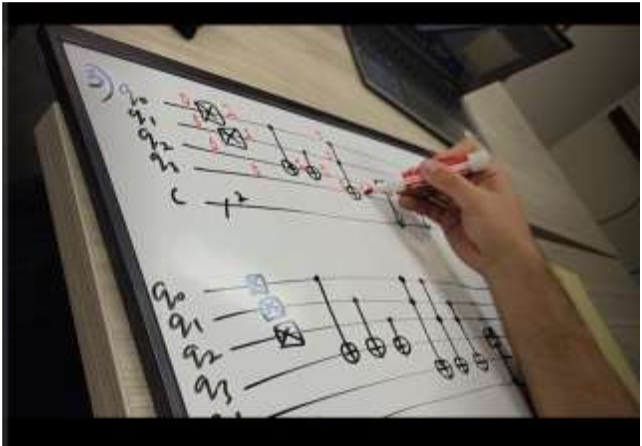

FIG. 6. Screenshot of a student showing step-by-step explanation of a quantum circuit

### B. Difficulties

The following are the reported difficulties common to 60% and above of our interviewed students.

- **Dirac operations:** From our results of the analysis, we observed that less than 40% of the students feel confident and comfortable working with Dirac notation, they prefer matrix method instead. In some computations Dirac notation is more beneficial than other notations, so it is useful for students to have this skill.

- **Interview excerpt:**
**Situation:** How were you able to know how the gates affect the qubit states?
**Response:** *Whenever I do these calculations, I always work in matrix, not Dirac.*
**Another response:** *I like to think about these things in terms of a matrix compared to using Dirac notations.*

- **Phase kickback:** This was one major difficulty common to almost all the interviewees. They could not explain what it really means to have a phase kickback. Some of the students had no idea about what phase kickback means while some could explain on a quantum circuit diagram how a controlled qubit changes phase while the target qubit remains unchanged based on the state of the target qubit. (Fig. 7)

Phase kickback is an important concept used in quantum algorithms. The overall effect is to add a phase shift to the control qubit, leaving the target qubit unchanged when a controlled gate is applied. This may mean we should introduce phase kickback in a different way or at a different point in the curriculum. Phase kickback provides a framework to understand many of the famous quantum algorithms that we have today, such as Shor's algorithm, Deutsch algorithm, and Simon's algorithm.

**Interview excerpt:**
**Situation:** What do you understand as phase kickback? (after explaining some circuit diagrams with the CNOT gate related to phase kickback)
**Response:** *I guess the CNOT does nothing on the qubit since the control is in a superposition state and not state one*

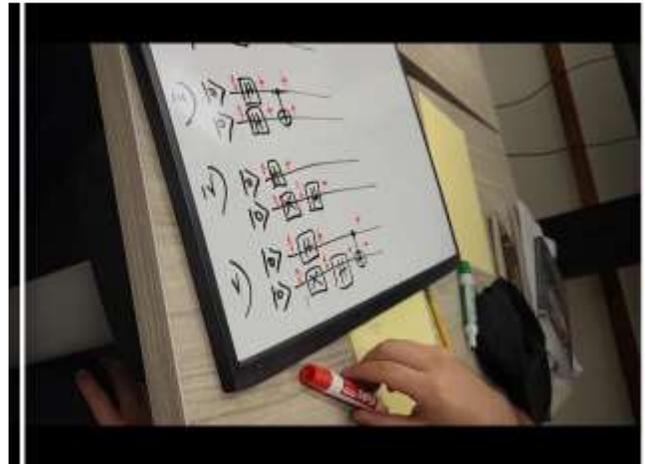

FIG. 7. *Screenshot of a student work showing his difficulty with phase kickback*

- **Rotation operators and Toffoli gate:** More that 60% of the interviewees could not tell how the rotation operators and Toffoli gate manipulate qubit states. The major difficulty was with representing the rotation operators in matrix and Dirac forms, which is important for computing the operations of these gates on qubits.

In our case, these gates were used less frequently in the course and we should possibly pay more attention to these gates in the future.

**Interview excerpt:**
**Situation:** How about these other gates, how do they manipulate the qubit state? (Fig. 8)
**Response:** *I am not sure how these other gates operate on qubits*
**Another response:** *hmmm I'm not gonna lie I actually don't know how the S gate works*

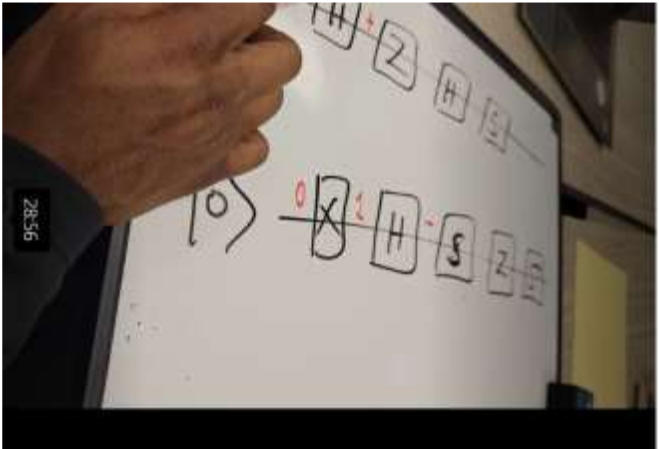

FIG. 8. Quantum circuit diagrams showing student having difficulty with some quantum gates

- ***Notations across various texts and online resources*:** Notation is a big confusion and difficulty for the students as reported by them. We heard such statements as '*using various texts can be confusing*.' These confusing notations have made it difficult for students to balance their learning from various texts and online resources.

## VI. IMPLEMETATION

We brainstormed ideas for ways to address some of these students' difficulties. One idea that we are working on is the development of multiple representation learning and homework materials [28] (a sample question is shown in Fig. 9). In the multiple representation learning materials, students would be made to answer a single question using different methods (matrix method, Dirac method, Bloch sphere, step-by-step circuit method, and conceptual explanation), i.e answering one question using several approaches, in order to master the different methods and understand how the methods are related for solving problems

## VII. CONCLUSIONS

We have researched students' strengths and difficulties in an Introductory Quantum Computing course. Even though this is a case study of five students, we expect that the results will be of some benefit to Quantum Computing instructors and Physics Education Researchers in the design of instructional materials and pedagogical approaches. This research also sets the stage for future larger studies of students' understanding of Quantum Computing topics. We have also presented some of our ideas for future materials development.

To conclude, this research is a part of ongoing pedagogical research and curriculum development. We will be conducting this qualitative study again to investigate the effectiveness of the research-based materials we develop and look at other issues, such as quantum mechanics pre-knowledge.

## Multiple representations sample question

(a.) Given the quantum circuit below, find the final state (without measurement) of the single qubit after all the gates operate on it. Answer this question using each of the methods below.

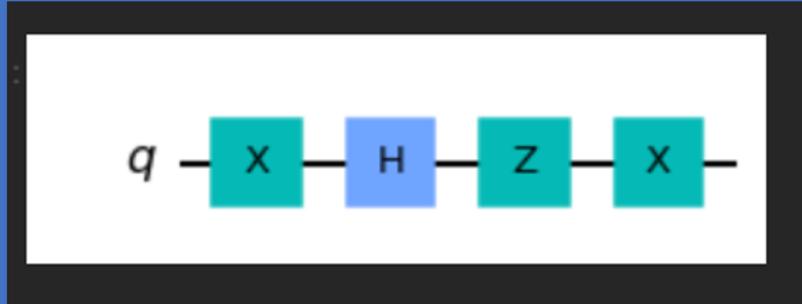

i. Matrix method
ii. Dirac method
iii. Bloch Sphere
iv. Step-by-step walk through the circuit (explain in your words how each gate manipulates the state of the qubit)

(b.) Write a quantum algorithm in python to represent the quantum circuit in (a.)

FIG. 9. Multiple representations sample question